\documentclass[reprint,pra,twocolumn,a4paper,superscriptaddress,aps]{revtex4}

\usepackage{graphicx}
\usepackage{dcolumn}
\usepackage{amsmath,amssymb,graphicx,color,dsfont}
\usepackage{mathtools}
\usepackage{pdfpages}
\usepackage{bm}
\allowdisplaybreaks
\usepackage{tabularx}
\pdfoutput=1
\usepackage{relsize}
\usepackage[toc]{appendix}
\usepackage{upgreek}
\usepackage{booktabs} 
\usepackage{multirow}
\usepackage{soul} 
\usepackage{microtype}

\newcommand{\T}{\textrm{T}/\sqrt{\textrm{Hz}}}
\newcommand*{\super}[1]{^{\raisebox{-4pt}{\tiny \rm{#1}}}} 
\newcommand*{\superl}[1]{^{\raisebox{-4pt}{\scriptsize ${#1}$}}}
\usepackage{xparse}
\ExplSyntaxOn
\NewDocumentCommand{\mref}{m}{\quinn_mref:n {#1}}
\seq_new:N \l_quinn_mref_seq
\cs_new:Npn \quinn_mref:n #1
 {
  \seq_set_split:Nnn \l_quinn_mref_seq { , } { #1 }
  \seq_pop_right:NN \l_quinn_mref_seq \l_tmpa_tl
  ( 
  \seq_map_inline:Nn \l_quinn_mref_seq
    { \ref{##1},\nobreakspace } 
  \exp_args:NV \ref \l_tmpa_tl 
  ) 
 }
\ExplSyntaxOff

\begin{document}

\title{
Modelling of cavity optomechanical magnetometers
}

\author{Yimin Yu}
\affiliation{ARC Centre for Engineered Quantum Systems, School of Mathematics and Physics, The University of Queensland, Brisbane, Queensland 4072, Australia}

\author{Stefan Forstner}
\affiliation{ARC Centre for Engineered Quantum Systems, School of Mathematics and Physics, The University of Queensland, Brisbane, Queensland 4072, Australia}

\author{Halina Rubinsztein-Dunlop}
\affiliation{ARC Centre for Engineered Quantum Systems, School of Mathematics and Physics, The University of Queensland, Brisbane, Queensland 4072, Australia}

\author{Warwick P. Bowen}
\email[]{w.bowen@uq.edu.au}
\affiliation{ARC Centre for Engineered Quantum Systems, School of Mathematics and Physics, The University of Queensland, Brisbane, Queensland 4072, Australia}
\date{\today}

\begin{abstract}
Cavity optomechanical magnetic field sensors, constructed by coupling a magnetostrictive material to a micro-toroidal optical cavity, act as ultra-sensitive room temperature magnetometers with tens of micrometre size and broad bandwidth, combined with a simple operating scheme. Here, we develop a general recipe for predicting the field sensitivity of these devices. Several geometries are analysed, with a highest predicted sensitivity of 180~p$\T$ at 28~$\upmu$m resolution limited by thermal noise in good agreement with previous experimental observations. Furthermore, by adjusting the composition of the magnetostrictive material and its annealing process, a sensitivity as good as 20~p$\T$ may be possible at the same resolution. This method paves a way for future design of magnetostrictive material based optomechanical magnetometers, possibly allowing both scalar and vectorial magnetometers.
\end{abstract}

\maketitle

\section{Introduction}
Magnetometers with high spatial resolution are required for many applications such as magnetoencephalography \cite{Encephal1993}, measurements of topological spin configurations \cite{Poggio2015} and nuclear magnetic resonance spectroscopy to identify chemical composition, molecular structure and \mbox{dynamics \cite{NMR2008}}.
Optical readout of magnetometers can offer high sensitivity for a given resolution, while being well decoupled from the magnetic signal. Among optical magnetometers, an ensemble of nitrogen-vacancy (NV) centres with a volume size of 8.5$\times 10^5~\upmu \rm m^3$ pushes the sensitivity down to 1~p$\T$ \cite{Wolf2015}. However, NV magnetometry generally requires high optical power for excitation (\mbox{e.g., 400~mW} in Ref.~\cite{Wolf2015}), as well as complicated microwave decoupling sequences in NMR spectroscopy, and is limited by the sample fabrication reproducibility \cite{Springer2017}.  A magnetometer based on micro-sized Bose--Einstein condensates has a volume of 90~$\upmu$m$^3$, but its quantum-enhanced sensitivity is limited to 1.86~n$\T$ \cite{Muessel2014}. It is crucial yet challenging to reduce the size of magnetometers while maintaining  competitive sensitivities. 

\begin{figure}[!ht]
\begin{center}
\includegraphics[width=0.9\columnwidth]{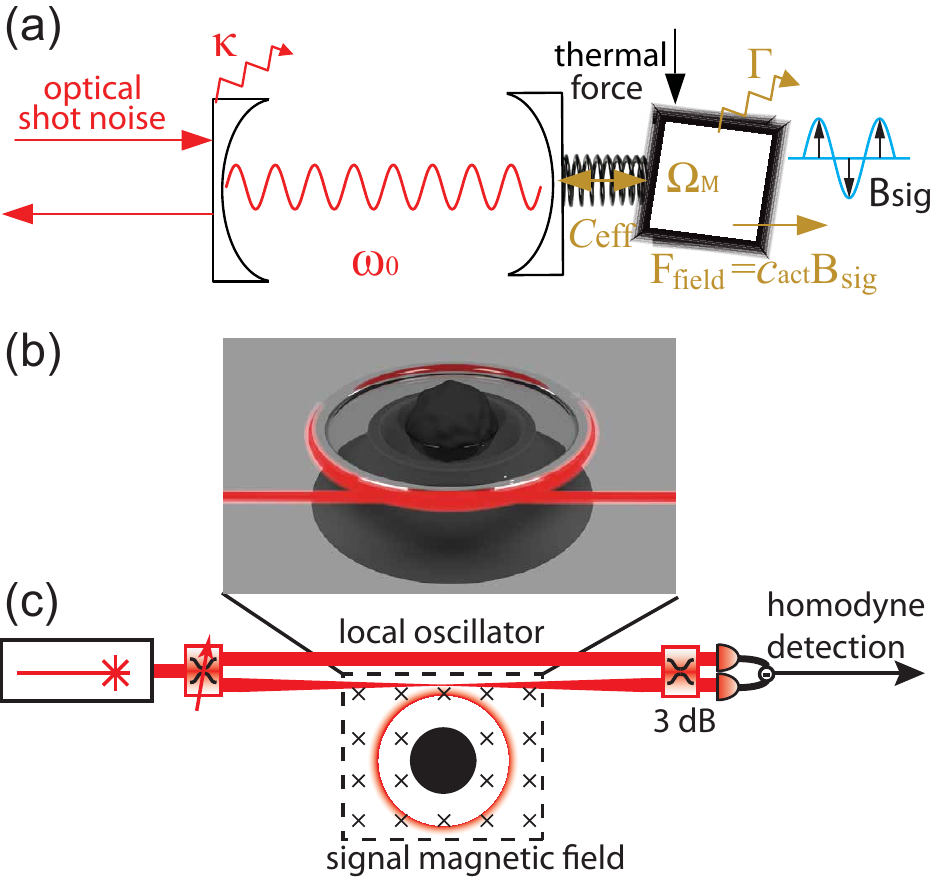}
\end{center}
\vspace{-3mm}\caption{
Concept of an optomechanical magnetometer. (\textbf{a}) illustration via a Fabry--P\'erot type optical resonator. The coupling of magnetostrictive material to an optical cavity is quantified by the effective cooperativity $C_{\rm eff}$. The magnetostrictive material converts a magnetic field to a force $F_{\rm field}=c_{act} B_{\rm sig}$ with $B_{\rm sig}$ being an oscillating magnetic field. Thermal force and optical shot noise act as noise terms. \mbox{$\kappa$ (rad$\cdot$s$^{-1}$)}, $\Gamma$ (rad$\cdot$s$^{-1}$), $\omega_0$ (rad$\cdot$s$^{-1}$), and $\Omega_{\rm M}$ (rad$\cdot$s$^{-1}$) are optical and mechanical decay rate, optical and mechanical resonance frequency, respectively;  (\textbf{b}) sketch of a magnetometer with micro-toroidal structure coupled to a tapered optical fibre; (\textbf{c}) homodyne detection scheme. The signal arm couples a coherent light source in and out from a magnetometer via a tapered optical fibre through an evanescent optical field, and is mixed with a strong reference beam (local oscillator field) by a 3~dB coupler. \mbox{The magnetometer} is embedded in the signal magnetic field.}
\label{Fig:setup}
\end{figure}

Among various types of magnetometers, optomechanical magnetometers \cite{Forstner2012,Forstner2014} reach sensitivities in the high p$\T$ range at room temperature with sizes of tens of micrometres, comparable to the best cryogenic SQUID-magnetometer of the same size \cite{Kirtley1994}. The principle of an optomechanical magnetometer is illustrated in {Figure}~\ref{Fig:setup}a. A magnetostrictive material converts the magnetic field to a force as a result of mechanical deformation. The magnetostrictive response has a nonlinear component, a property that has been utilised in previous work to mix low frequency magnetic fields up to megahertz frequencies and therefore evade low frequency noise~\cite{Forstner2014}. However, in~general, it~is far smaller than the linear component, so that the force may be well approximated by \linebreak$F_{\rm field}=c_{act}B_{\rm sig}$, where $c_{act}$ (N/T) is the actuation parameter and $B_{\rm sig}$~(T) is the magnetic field to be measured. \mbox{The amplitude} of the mechanical response to this force is greatly enhanced when the magnetostrictive material is driven resonantly at its mechanical eigenfrequency by a modulated magnetic field. The mechanical response changes the path length of the optical cavity to which the magnetostrictive material is attached, allowing the magnetic field to be read out optically from the shift of the optical resonance \cite{Bowen2015}. While significant successes have been achieved in experimental demonstrations of optomechanical magnetometers \cite{Forstner2012,Forstner2014}, modelling and sensitivity-prediction for these devices have been somewhat \textit{\mbox{ad hoc}}~\cite{Forstner2012P,Forstner2012SPIE}. Better modelling techniques are needed to both enhance understanding of previous experimental results and for design of future magnetometers.

\textls[-15]{In this work, we present a model of magnetostrictive magnetometers that accounts for arbitrary mechanical mode shape and device geometry. We modify the elastic wave equation, which describes the small-amplitude motion of elastic materials, by including magnetostrictive stress. This modified elastic wave equation is then numerically solved by finite element analysis (using COMSOL Multiphysics). Magnetomechanical overlap, describing the overlap between the magnetostrictive deformation induced by the signal magnetic field and the excited mechanical eigenmode, is intrinsically included in the matrix form of the modified elastic wave equation, with each matrix element containing directional information. Mechanical properties are extracted from the solution to the modified elastic wave equation from COMSOL to be further combined with optomechanical analysis \cite{Bowen2015} to predict the sensitivity of a magnetometer for a given geometry.}

We apply this analysis to study the effect of the position of the magnetostrictive material on the sensitivity of devices similar to those reported in Ref.~\cite{Forstner2012}. Using the piezomagnetic constant measured from a rod of the magnetostrictive material Terfenol-D \cite{Claeyssen1991}, we model a magnetometer design, where the Terfenol-D is deposited directly on top of a standard silica toroid. From there, we employ single mode analysis (Appendix~\ref{app:single}) and discover that a bimetallic-stripe-like bending effect, similar to the bimetallic bending effect in a cantilever \cite{bimetallic2007}, greatly enhances the sensitivity when the magnetostrictive material is positioned off-centre. Optimisation of this effect may allow substantial improvements in sensitivity in future devices. Furthermore, we investigate the sensitivity achievable from a device comprised of a toroidal structure with a centre hole that is filled with the Terfenol-D, as studied experimentally in Ref.~\cite{Forstner2014} and sketched in Figure~\ref{Fig:setup}b . We predict a peak sensitivity of {180}~p$\T$ over a broad spectrum by using multi-mode analysis under optimised operational conditions, {in~good agreement with} current experimental observations. 

This numerical model allows specification of the orientation of a sample to maximally enhance the magnetomechanical overlap, thus amplifying the detected magnetic field signal, as well as characterization of the magnetomechanical overlap in response to the variation of the magnetic field direction. This is crucial to vectorial magnetometers that measure not only the intensity but also the direction of the magnetic field.

\section{Concept of Optomechanical Magnetometry}
Optomechanical magnetometry can be schematically explained via the example of a Fabry--P\'erot optical resonator coupled to a spring-mass mechanical oscillator as depicted in Figure~\ref{Fig:setup}a. An applied magnetic field $B_{\rm sig}$ causes a deformation to a magnetostrictive material attached to the mechanical oscillator (see Appendix~\ref{app:MagneticField} for details of how this field is generated in COMSOL). This induces a \linebreak$F_{\rm field}=c_{act}B_{\rm sig}$ on one movable end mirror of the optical resonator, changing the optical path length and thus the optical resonance frequency. The shift in the optical resonance frequency is therefore proportional to the applied magnetic field. The transduction from magnetic field to mechanical motion is determined by the actuation parameter $c_{act}$ depending on magnetomechanical overlap and magnetostrictive coefficient. { The magnetic field signal encoded on the motion of the mechanical element is read out by optically probing the the optical resonance frequency. This can be achieved with high precision by coupling a coherent optical field into the cavity, collecting the output field, and measuring the change in its amplitude or phase due to the modulation of the optical resonance frequency. For instance, directly detecting the output field, as in  several reported experiments~\mbox{\cite{Forstner2012, Forstner2014}}, measures changes to the amplitude of the output optical field and enables simple operation. Alternatively, a homodyne scheme can be used, allowing an arbitrary quadrature of the optical field to be accessed as shown in Figure~\ref{Fig:setup}c. Here, the output field is interfered with a bright local oscillator field prior to detection. The transduction from mechanical displacement to optical signal can be quantified by the effective cooperativity $C_{\rm eff}$~\cite{Bowen2015}.
}

The magnetic field sensitivity is limited by noise consisting primarily of thermal force and shot noise on the optical field. Thermal noise is explained by the equipartition theorem, which states that each mechanical degree of freedom of an object has a mean energy of k$_B T/2$ (k$_B$ is the Boltzmann constant and $T$ is the temperature). This energy excites incoherent mechanical vibration near  mechanical eigenfrequencies. The bandwidth of the magnetometer depends on the visibility of the thermal noise  over the optical shot noise. For the case of a single mechanical resonance, the sensitivity is flat over the frequency range where thermal noise dominates shot noise, and~degrades outside of this region. Consequently, in this case, the bandwidth is given simply by the thermal-noise-dominant frequency band, which is typically on the order of a few megahertz \cite{Schliesser2008NJP}. \mbox{The case} of multiple mechanical modes is more complex due to variations in actuation constants, effective cooperativities and mechanical parameters, and due to interferences in the coherent response of the mechanical~modes. 

\begin{figure}[!ht]
\begin{center}
\includegraphics[width=0.9 \columnwidth]{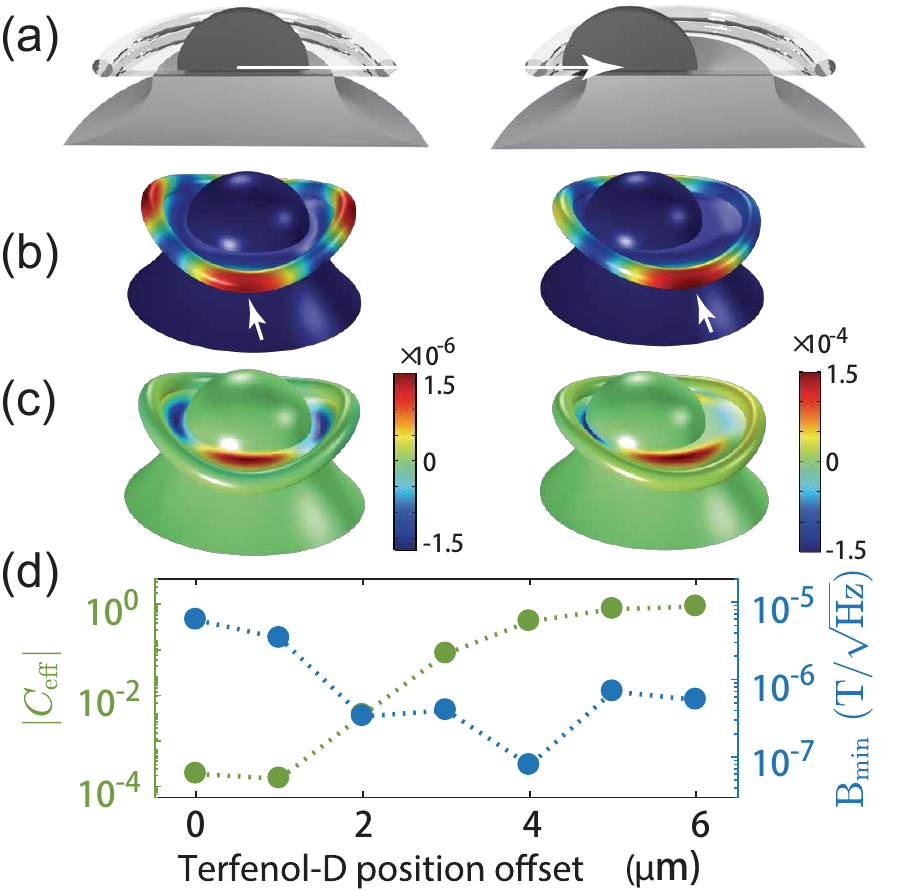}
\end{center}
\vspace{-4mm}\caption{
(\textbf{a}) sketch of the position offset of the magnetostrictive material of the first experimentally realized optomechanical magnetometers \cite{Forstner2012}; (\textbf{b}) a second order crown mode without (left) and with 4~$\upmu$m (right) Terfenol-D position offset. Arrows show the positions with maximum displacement; \mbox{(\textbf{c}) strain} of the magnetometer with centred (left) Terfenol-D, and with 4~$\upmu$m offset (right). Note that the colourmaps of the strain have different scales; (\textbf{d}) $\lvert C_{\rm eff}\rvert$ and sensitivity as a function of the position of the Terfenol-D.
}
\label{Fig:singlemodes}
\end{figure}

In this paper, as a test geometry for our model, we choose optomechanical magnetometers of the form reported in Refs.~\cite{Forstner2012, Forstner2014}. They utilise a silica microtoroid as the optical resonator. \mbox{The magnetostrictive} material is embedded in or deposited onto the microtoroid as sketched in {Figures}~\ref{Fig:setup}b and~\ref{Fig:singlemodes}a,  respectively. Combined, the silica microtoroid, the magnetostrictive material and the silicon pedestal serve as the mechanical oscillator. Using a tapered optical fibre placed next to the toroid, the optical field can be coupled in and out of the microtoroid through an evanescent optical field. This optomechanical magnetometry platform offers a simple operational scheme and low energy consumption with state-of-the-art field sensitivity for a micro-magnetometer. 

\section{Numerical Methods \label{chNm}}

The primary objective of this work is to develop a versatile technique to numerically obtain a meaningful estimation of the magnetic field sensitivity for a wide range of sensor geometries. \mbox{We consider} the case of phase quadrature detection in a homodyne scheme and on-resonance optical probing of the cavity resonance, which maximises the signal-to-noise. { We note, however, that simpler \mbox{direct} detection with off-resonance probing and an optimal detuning of  $\frac{\kappa}{2\sqrt{3}}$, where $\kappa$ is the optical cavity linewidth, only degrades the sensitivity by a factor of $\frac{8}{3^{3/2}} \sim 1.5$.  

The sensitivity as a function of magnetic field frequency $\Omega$ can be determined from the finite-time sensor power spectrum $S(\Omega)$, which can be separated into a stochastic noise term $S_{\rm noise} (\Omega)$ and coherent signal term $S_{\rm signal} (\Omega)$ as

\begin{equation}
S(\Omega)=\tau^{-1}\langle i^*(\Omega)i(\Omega)\rangle=S_{\rm signal}(\Omega)+S_{\rm noise}(\Omega),
\end{equation}
where $i$ is the photocurrent, normalised so that the optical shot noise contribution to $S_{\rm noise} (\Omega)$ is equal to $1/2$ \cite{Bowen2015}, and $\tau$ is the measurement time. At frequencies $\Omega \gg 2\pi/\tau$ and considering $j$ mechanical modes, $S_{\rm noise}(\Omega)$ is given by \cite{Bowen2015}

\begin{equation}
\begin{aligned}
S_{\rm noise}(\Omega)&=\frac{1}{2}+\mathlarger{\sum}_{j} 8\eta\Gamma_j^2\lvert C_{{\rm eff},j}(\Omega)\rvert \lvert\chi_j(\Omega)\rvert^2\Big[\frac{k_BT}{\hbar\Omega_{{\rm M},j}}\\&+\lvert C_{{\rm eff},j}(\Omega)\rvert \Big],
\label{S0}
\end{aligned}
\end{equation}
\textls[-15]{where the first term is the optical shot noise and the second term constitutes the combination of mechanical thermal noise and quantum back-action noise. The detection efficiency $\eta$, consisting of the loss in the fibre-device coupling and detection process, is ideally taken to be 1 in the model. However, \mbox{in the} non-back-action dominated regime relevant here, reductions in efficiency can be exactly modelled by a proportionate decrease in the optomechanical \mbox{cooperativity}. $\Gamma_j$ is the mechanical decay rate of mode $j$ and $C_{{\rm eff},j}$ is its effective cooperativity, which depends on the input laser power used, the decay rate of the optical field and mechanical excitation, and the radiation pressure coupling rate between them. The mechanical susceptibility of mode $j$ is defined as $\chi_j(\Omega)\equiv\Omega_{{\rm M},j}/(-\Omega^2-i\Omega\Gamma_j +\Omega_{{\rm M},j}^2)$, with $\Omega_{{\rm M},j}$ its mechanical resonance frequency. $k_B T/\hbar\Omega_{{\rm M},j}$ is the number of phonons thermally excited at room temperature, with $\hbar$ being the reduced Planck constant. The mechanical motion induced by an alternating-current (AC) magnetic field is quantified by the finite-time power spectrum $S_{\rm signal}(\Omega)$. This is calculated by replacing the thermal environment forcing $F_{\rm th}$ in the input momentum fluctuation \mbox{$P_{in}=x_{\rm zpf}F_{\rm th}/\hbar\sqrt{\Gamma}$ \cite{Bowen2015}}, which leads to Equation~\eqref{S0}, with a coherent sinusoidal driving force $F_{\rm field}(t)=c_{act} B(t)_{\rm sig}$ at frequency $\Omega$, and neglecting the incoherent noise terms (laser shot noise in amplitude and phase quadrature). This results in the expression}

\begin{equation}
\begin{aligned}
 S_{\rm signal}(B_{\rm sig},\Omega)=& 16\tau\pi\eta \Bigg\lvert\mathlarger{\sum}_{j} \Gamma \sqrt{\lvert C_{{\rm eff},j}(\Omega)\rvert} \chi_j(\Omega)\\
 &\cdot\frac{c_{{act},j}B_{\rm signal,rms}}{\sqrt{4 m_{{\rm eff},j}\Omega_{{\rm M},j}\hbar\Gamma_j}}\Bigg\rvert^2,
\label{SBB0}
\end{aligned}
\end{equation}
where $B_{\rm signal,rms}$ is the root-mean-square amplitude of $B_{\rm sig}(t)$, $m_{{\rm eff},j}$ is the effective mass of mode $j$, and~$c_{{act},j}$ is the actuation constant associated with that mode. This finite-time power spectrum takes into account mechanical interference, as experimentally observed, for example in optoelectromechanical systems coherently driven by an electric field \cite{Lee2010}.  

The frequency dependent signal-to-noise ratio (SNR) of the magnetic field measurement is given simply by

\begin{equation}
\text{SNR}= \frac{S_{\rm signal}(B_{\rm sig},\Omega)}{S_{\rm noise}}.
\end{equation}

The minimum detectable field in the measurement time $\tau$ is defined as the field that produces a signal-to-noise ratio SNR of one, i.e., $B_{{\rm min},\tau} = B_{\rm sig}({\rm SNR}=1)$. It should be noted that the stochastic noise power spectral density $S(\Omega)_{\rm noise}$ of Equation~(\ref{S0}) is independent of integration time, whereas the integral of a coherent band-limited signal power spectrum, as described by $S_{\rm signal}(\Omega)$ in Equation~(\ref{SBB0}), increases linearly with time. Consequently, $B_{{\rm min},\tau}$ improves with measurement time as $\tau^{-1/2}$. To obtain a minimum detectable field in the conventional units of Tesla per root Hertz, independent of time, we multiply through by $\tau^{1/2}$ with the result

\begin{align}
B_{\rm min}(\Omega)=B_{{\rm min},\tau} (\Omega)\times \tau^{1/2} = B_{\rm sig}\sqrt{\frac{S(\Omega) \times \tau}{ S_{\rm signal}(B_{\rm sig},\Omega)}}.
\label{BminM}
\end{align}

To determine the minimum detectable field via finite element simulations, we use COMSOL Multiphysics. Simulations detailed in the appendices allow us to extract each of the parameters in Equations~(\ref{S0})~and~(\ref{SBB0}) and therefore predict the sensitivity. These simulations involve both mechanical eigenmode solving to determine the resonance frequency, effective mass and effective cooperativity of each mechanical mode of a given device geometry; and magnetic field driving to determine the coherent response of the mechanical modes to a magnetic field and the interferences between them. The approach is briefly sketched in what follows.

\textls[-15]{The spatio-temporal mechanical modeshape is described by a separable function \mbox{$\bm u(\bm r,t)=\bm \Psi(\bm r)x(t)$}. The effective mass $m_{{\rm eff},j}$ for one mechanical resonance at an eigenfrequency $\Omega_{{\rm M},j}$  is calculated from the maximum physical displacement $max_{\bm r}[|\bm \Psi(\bm r)|]$ as $m_{{\rm eff},j}=\int_V \rho_n \bm |\Psi(\bm r)|^2 dV$ \cite{Davis2013}, with normalization 
{$max_{\bm r,t} [|\bm u({\bm r,t})|]=max_t[x(t)]$} and therefore $max_{\bm r} [|\bm \Psi(\bm r)|]^2 =1$. $\rho_n$ is the density of the material and the subscript $n$ denotes different parts of the device (for instance, silica for the optical resonator and Terfenol-D for the transducing medium). Note that, while this definition of effective mass is the convention for microelectromechanical systems, an alternative definition---where the effective mass is defined with respect to the optical path length---is commonly used in the optomechanical community \cite{Pina1999}. This choice of convention has no effect on the ultimate predictions of our model.}

The magnetic field response  $S _{\rm signal}(\Omega)$ of the sensor is determined by the eigenmode-dependent actuation parameter $c_{act}$. For a single mechanical eigenmode, the equation of motion is

\begin{equation}
\ddot{x}(t)+\Gamma \dot{x}(t)+\Omega^2_{\rm M} x(t)=\frac{c_{{act}}B_{\rm sig}(t)}{m_{{\rm eff}}}. \label{asd}
\end{equation}

At the resonance frequency of each mechanical eigenmode, $c_{act}$ can be extracted as a fitting parameter
in the mechanical signal frequency response spectrum obtained from COMSOL. Taking the Fourier transform of  Equation~(\ref{asd}), we see that

\begin{equation}
 c_{{act}} = \frac{x(\Omega) \Omega_M m_{{\rm eff}}}{\chi(\Omega)  B_{\rm sig} (\Omega)}.
\end{equation}
This allows $c_{{act}}$ to be determined for each mechanical mode.}

Due to the magnetostrictive energy stored within compressed magnetostrictive materials, \mbox{the extraction} of $m_{\rm eff}$ and displacement from COMSOL for such materials requires modification of the elastic wave equation. To treat the magnetostrictive material in COMSOL,  we built upon a previously used method \cite{Terfenol,CLAEYSSEN1997,Kannan1997}, including the magnetic field in a driving stress $\bm \sigma_{\rm driv}$ and adding a damping stress $\bm \sigma_\Gamma$ to the elastic stress $\bm \sigma_{\rm ela},$ which describes the mechanical properties without driving force in the elastic wave equation \cite{LandauElas}, resulting in 

\begin{equation}
-\rho_n \Omega^2 \bm u  = \triangledown\cdot\hm(\bm \sigma_{\rm ela}+ \bm \sigma_{\rm driv}+\bm \sigma_{\Gamma}).
\label{EqMdamping}
\end{equation}

The modulated driving stress is linked to the magnetic field via the piezomagnetic constant \cite{CLAEYSSEN1997}, and a low value for the damping stress  $\sigma_{\Gamma}$ is chosen manually to avoid an artefactual infinity in the mechanical displacement at resonance (see Appendix \ref{app:ElasticW} for technical details). Simulations reveal that the influence of a particular value chosen for $\sigma_\Gamma$ on numerical results is negligible (Appendix \ref{app:cact}).

To obtain the value of effective cooperativity, we quantify the effectiveness of transduction of mechanical motion to measurable optical path length change as the geometrical factor, as

\begin{align}
\xi\equiv\frac{\delta L}{{\it max}[|\bm u|]},\label{xi}
\end{align}
where $\delta L$ is the change of the optical path length due to the mechanical displacement. The extraction of the value of $\xi$ from COMSOL is detailed in Appendix \ref{app:xi}. Within one mechanical mode, $\xi$ is directly linked to the effective cooperativity (Appendix \ref{app:fromxi}) by

\begin{equation}
\lvert  C_{\rm eff}(\Omega)\rvert=\frac{\xi^2}{m_{\rm eff}\Omega_{\rm M}\Gamma}\cdot \frac{8\eta_{esc}\hbar N_{in}\omega_0^2}{L^2\big( \kappa^2+4\Omega^2\big ) }\label{ceff-xi}
\end{equation}
where $\kappa$ is the optical decay rate, $\omega_0$ is the optical resonance frequency, $N_{in}$~(photons$\cdot$s$^{-1}$) is the input optical photon number  flux, $L$ is the optical path length, and $\eta_{esc}$ is the escape efficiency counting fibre-device coupling. The front part of the right hand side of Equation~\eqref{ceff-xi} is arranged  to be  mechanical mode dependent. The calculation of the magnetic field sensitivity from Equations~(\ref{S0}), (\ref{SBB0}) and (\ref{BminM}) can then be obtained based on the value of the geometrical factor $\xi$.

\section{Single Mechanical Mode Optomechanical Analysis}
\subsection{Bending Effect}

To verify the numerical model, we apply it to the first experimentally realized optomechanical magnetometer \cite{Forstner2012}. For simplicity, we begin the analysis considering only a single mechanical eigenmode (Appendix \ref{app:single}). The magnetometer as sketched in Figure~\ref{Fig:singlemodes}a consists of a silica micro-toroidal cavity with major radius of 33~$\upmu$m. The Terfenol-D is glued on top of the silica and is modelled as a semi-sphere with a transverse radius of 18.5~$\upmu$m and a height of 15~$\upmu$m. The optical quality factor $Q_o=\omega_o/\kappa$ is taken to be 2$\times10^7$ from the experiment. The mechanical quality factor $Q_{\rm M}=\Omega_{\rm M}/\Gamma$ is assumed to be 200 for all modes which is a simplification, but is roughly in line with the experimentally observed quality factors. A continuous input laser is locked to the optical cavity resonance in the homodyne detection scheme, and the input laser power ensures that on mechanical resonances thermal noise dominates over optical shot noise. 

From available optical microscopic images, it is not clear whether the Terfenol-D is centred on the toroid or not. Therefore, we sweep the position of the Terfenol-D from the centre. Without loss of generality, we analyse the magnetic response for a second order crown mode because this mode has been commonly observed in experiments \cite{Wilson2016,Schliesser2008NJP,Lee2010}. For the magnetometer with centred Terfenol-D, the effective motional mass is $m_{\rm eff}=3.9$ pg with eigenfrequency at 10.1 MHz. As the Terfenol-D is moved away from the centre as illustrated in Figure~\ref{Fig:singlemodes}a, the mechanical eigenmode changes (Figure~\ref{Fig:singlemodes}b).  Generally, the top of the Terfenol-D stretches more than the bottom part attached to a silica disk during a mechanical oscillation. This is also the case for silica where the top layer experiences the force from the Terfenol-D and the bottom layer is clamped to the silicon pedestal. Therefore, a bimetallic-like strain gradient is formed vertically. In the second order crown mode, as the major motion takes place at the silica layer instead of the Terfenol-D, the strain gradient can be viewed inside the silica disk at the edge of bottom Terfenol-D and top facet of silicon. With the centred Terfenol-D, the strain at the top layer of the silica is nearly two orders of magnitude smaller than that with 4~$\upmu$m Terfenol-D position offset as shown in the red areas in Figure~\ref{Fig:singlemodes}c. This local maximum strain leads to the maximum displacement of the device (pointed by the arrows on tori in Figure~\ref{Fig:singlemodes}b) in the radial direction. Figure~\ref{Fig:singlemodes}d shows the best sensitivity of 78~n$\T$, when driven by an in-plane magnetic field, takes place when the Terfenol-D offset is at 4~$\upmu$m, nearly two orders of magnitude better than that of Terfenol-D centred (the same order of magnitude difference as that of the strain). We therefore see that the position of the Terfenol-D on the silica layer has strong influence on the bimetallic-like strain effect, and consequently the sensitivity. Moreover, the effective cooperativity $\lvert C_{\rm eff}\rvert$ of the crown mode experiences four orders of magnitude enhancement with only a few micrometres Terfenol-D offset as plotted in Figure~\ref{Fig:singlemodes}d. $\lvert C_{\rm eff}\rvert$ is chosen for evaluating mechanical mode shape induced characteristics. Terfenol-D with offset breaks the axial symmetry of the crown mode, creating a first order circumference difference of the toroid as the mechanical mode oscillates, and thus improves the value of $\lvert C_{\rm eff}\rvert$.
  
The numerical results show that asymmetry and the bimetallic-like bending effect helps to enhance the sensitivity. With an optimal offset of Terfenol-D, low n$\T$ sensitivity is predicted, which is five~times better than the experimental result \cite{Forstner2012}. It is likely that the experimental results were degraded not only due to a lack of Terfenol-D offset, but also by the epoxy  used to fix the Terfenol-D on top of the toroid, reducing the expansion of the silica disk. 

\subsection{Effect of the Size of the Terfenol-D}
\begin{figure}[!ht]
\begin{center}
\includegraphics[width=0.9 \columnwidth]{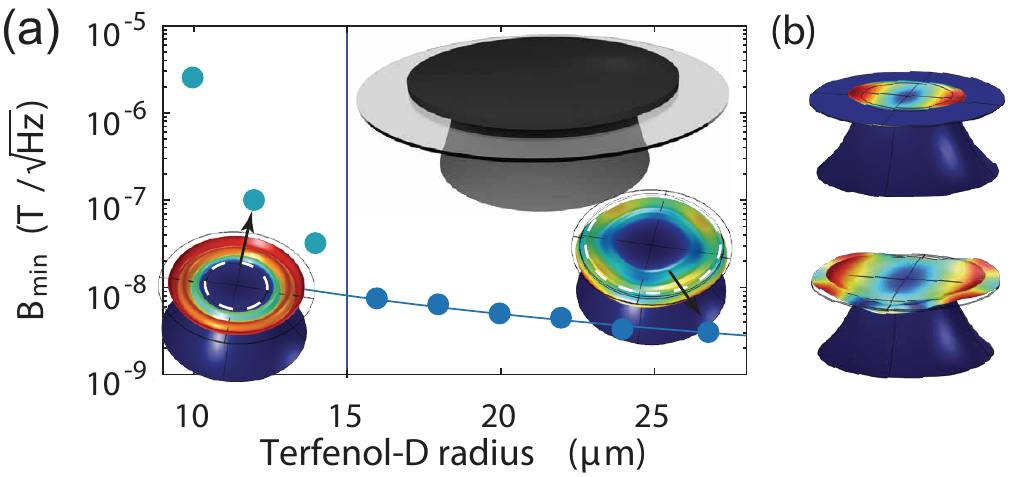}
\end{center}
\vspace{-4mm}\caption{
(\textbf{a}) sensitivity vs. Terfenol-D disk size for the first order radial breathing-type modes of a thin film structure. The silica disk dominates the mechanical eigenmodes when the Terfenol-D (highlighted with white dashed line) is smaller ({left})~than the 15~$\upmu$m radius top facet of silicon pedestal indicated by a vertical line. If the Terfenol-D is larger than the silicon facet, the mechanical motion is hybridized with the Terfenol-D mode  ({right)}.~A power-law fit is applied to the right side data. Insets are sketches of a thin film magnetometer and of two mechanical eigenmodes; (\textbf{b}) deformation profile induced by axial magnetic field driving for Terfenol-D smaller and larger than the pedestal top facet. 
}
\label{Fig:size}
\end{figure}

The single mechanical mode analysis is then applied to a proposed \cite{Forstner2012} thin disk structure: 1~$\upmu$m sputter coated Terfenol-D film  on top of a 400~nm-thick silica disk. Magnetometers with sputter coated Terfenol-D have the advantage of a reproducible fabrication process. The silica disk has a radius of 30~$\upmu$m and the pedestal has a top facet of 15 $\upmu$m (sketched in Figure~\ref{Fig:size}a inset top). The optical quality factor is taken to be $1\times 10^6$ \cite{Vahala2003}, a coherent laser source is again used to probe the system with zero detuning and measured via homodyne detection. The effective mass extracted from numerical simulation varies from 1~pg to 3.8~pg depending on the Terfenol-D size, for the radial breathing modes of the device. 

Figure~\ref{Fig:size}a shows the relation of sensitivity to  the size of the Terfenol-D for the first order radial breathing-type mode. The signal magnetic field drives the radial breathing mode in the axial direction to create a magnetic field induced deformation profile as shown in Figure~\ref{Fig:size}b. Unlike an isotropic magnetostrictive material breathing radially under axial magnetic field driving, \mbox{the spatial} profile from the non-isotropic Terfenol-D stretches only in one direction. When the size of the Terfenol-D is larger than the top facet of the silicon pedestal, the silica disk is also significantly affected by the motion from the Terfenol-D. The part of Terfenol-D inside the top pedestal facet (Terfenol-D is highlighted with the white dashed line in the mechanical eigenmode simulation in Figure~\ref{Fig:size}a inset) is motionless because it is obstructed by the silicon pedestal. When the rim of the Terfenol-D reaches outside the top pedestal facet, the device mechanical motion is hybridized with mechanical modes of the Terfenol-D. This leads to a bi-metallic-strip-like effect close to the edge of the top facet of the silicon pedestal across the silica layer, increasing the silica displacement and thus allowing for better sensitivity than in the cases where the Terfenol-D is confined inside the silicon pedestal. Generally, \mbox{the sensitivity} scales with the size of the motional part of the Terfenol-D. A sensitivity of 2.9~n$\T$ is predicted when the diametre of the Terfenol-D disk covers more than 2/3 of the silica disk in Figure~\ref{Fig:size}a. A power-law  fit ($y(x)=a\cdot x^b$ with fitting results of $a=7.9\times10^{-7}$ and $b=-1.7$) is applied to the data with the Terfenol-D radius larger than that of the pedestal, predicting a 300~$\upmu$m radius of Terfenol-D may lead to 50~p$\T$ sensitivity. To achieve better sensitivity, the size of the Terfenol-D must be larger than the pedestal so as to have large portion of motional Terfenol-D and large bi-metallic-strip-like bending effect, which could be realised by decreasing the size of the silicon pedestal and by increasing the size of the Terfenol-D.

\section{Multi-Mode Analysis}
\label{multi-simu}
Single mode analysis is limited, in that it only correctly predicts the performance of devices over frequency ranges where only one mechanical mode contributes significantly to the dynamics. In reality, this is rarely the case, and often there is a dense spectrum of mechanical modes (\mbox{see e.g., Ref.~\cite{Forstner2012,Forstner2014}}). \mbox{To extend} our analysis to such situations, we use multi-mode analysis from Section~\ref{chNm}. \mbox{We first} examine the limitations of the single mode analysis and then predict an optimal driving direction of the magnetic field leading to a best predicted sensitivity of an ensemble of mechanical eigenmodes.

\begin{figure*}[!hptb]
\begin{center}
\includegraphics[width=1.8\columnwidth]{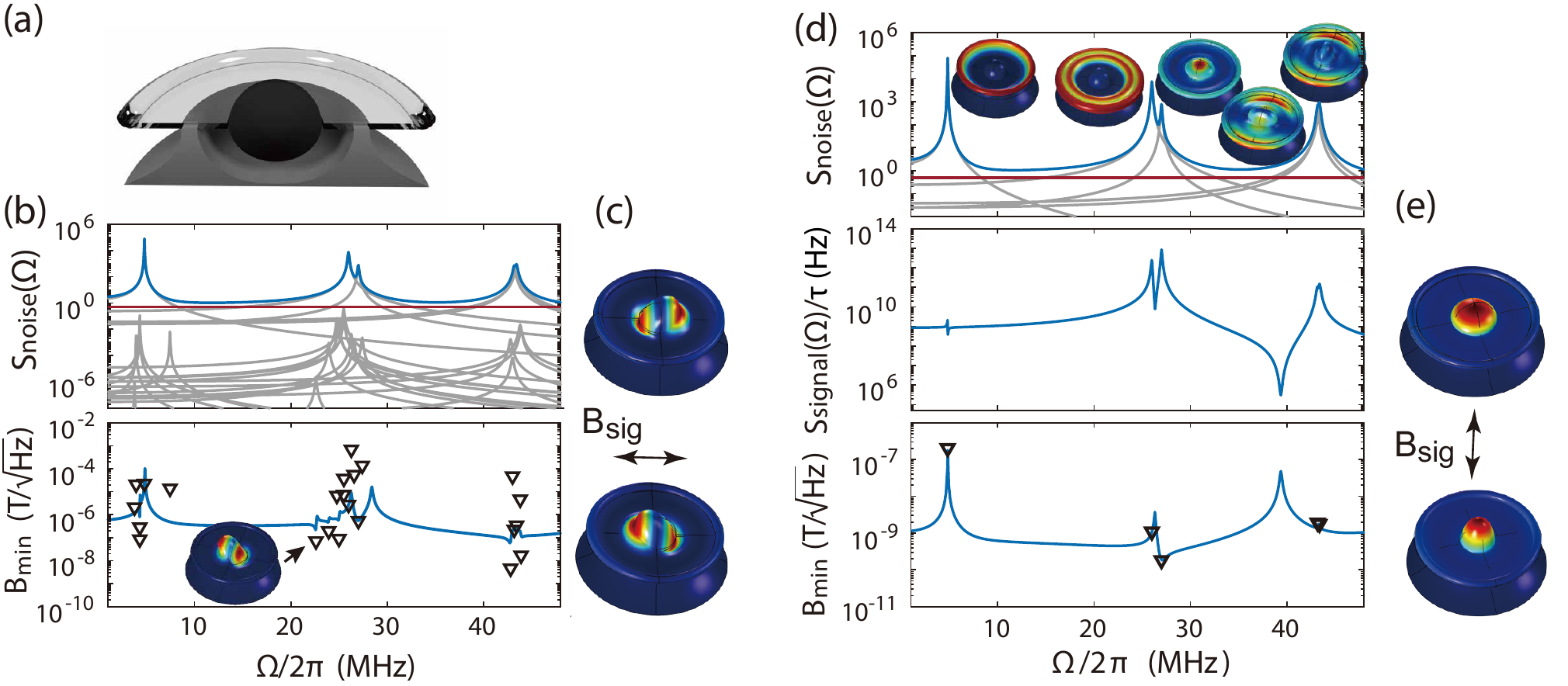}
\end{center}
\vspace{-3mm}\caption{Multi-mode analysis with a device reported in Ref.~\cite{Forstner2014}. (\textbf{a}) cross-sectional view of the optomechanical magnetometer; (\textbf{b}) top: the power spectral density $S_{\rm noise}(\Omega)$ ({blue}) is the sum of individual thermal Brownian motion peaks ({grey}) and coherent laser shot noise on the optical phase quadrature ({red}); bottom: minimum detectable magnetic field from multi-mode ({blue}) and single mode (black triangles) analysis driven by in-plane magnetic field. The inset is the mechanical mode with the highest $c_{act}$ at $\Omega_{\rm M}/2\pi$ = 23~MHz; (\textbf{c}) deformation profile induced by in-plane magnetic field far away from mechanical resonance frequencies; (\textbf{d}) top: the power spectral density $S_{\rm noise}(\Omega)$ of the radial-breathing-like mechanical modes. The insets show the mechanical eigenmodes corresponding to each resolved thermal Brownian motion peaks; middle: the magnetic field response $S_{\rm signal}(\Omega)/\tau$ to the axial magnetic field driving; bottom: the sensitivity spectrum from multi-mode ({blue}) and single mode (black triangles) analysis driven by the axial magnetic field; (\textbf{e}) deformation profile induced by axial magnetic field far away from mechanical resonance frequencies.
}
\label{Fig:hv}
\end{figure*}

We examine the limitations of single mode analysis by considering the magnetometer design reported in Ref.~\cite{Forstner2014}. This type of magnetometer has a hole of 14~$\upmu$m radius in the middle of a silica toroid, which has a 45~$\upmu$m major radius. A cross-sectional view is shown in Figure~\ref{Fig:hv}a, where the outer silicon undercut is 15~$\upmu$m. The Terfenol-D is modelled as an ellipsoid having the same transverse radius as the silica hole and an axial radius of 16~$\upmu$m. Mechanical modes with resonant frequencies up to 45~MHz are selectively driven with the in-plane $ B_{\rm sig}$ in accordance with the experimental conditions of Ref.~\cite{Forstner2014}. Three windows ($\sim$7~MHz, $\sim$26~MHz and $\sim$43~MHz) of interest are selected. Mechanical modes in between are not taken into consideration due to their small optomechanical coupling resulting from their symmetrical mode shapes. The power spectral density $S_{\rm noise}(\Omega)$ and magnetic field sensitivity spectrum in Figure~\ref{Fig:hv}b are obtained, again choosing $Q_{\rm M}=200$ for all modes, and setting $Q_o=2\times 10^6$ and a coherent laser with power of 1~$\upmu$W at 1550~nm in an on-resonance homodyne detection scheme. With these parameters, the sensor noise floor is dominated by mechanical thermal noise close to the mechanical resonance frequencies, and optical shot noise at other frequencies (Figure~\ref{Fig:hv}b top). \mbox{A single} mechanical mode at $\Omega_{\rm M}/2\pi$=23~MHz has the largest actuation parameter (see Appendix~\ref{app:cact} for $c_{act}$ spectrum) due to a relatively large spatial mode overlap  between the mechanical eigenmode (\mbox{Figure~\ref{Fig:hv}b inset}) and the magnetic field induced deformation profile (Figure~\ref{Fig:hv}c) compared with other modes. However, this particular mode has a very weak optomechanical coupling when the device is modelled uniformly and axial-symmetrically. This prevents the mode from being optically resolved from the thermal noise of others, causing a large difference of the magnetic field sensitivity between the single mode and multi-mode analysis, as shown in triangles and lines in Figure~\ref{Fig:hv}b bottom, respectively.

To achieve better sensitivity, the direction of the driving magnetic field needs to be optimised. The mechanical mode under magnetic field driving should have both relative large optomechanical coupling and relative good magnetomechanical overlap compared to other modes. As might be expected, and is shown in Figure~\ref{Fig:hv}b, top modes with radial-breathing-like motion (Figure~\ref{Fig:hv}d top insets show the eigenmodes) offer the largest optomechanical coupling. These  mechanical modes are at 4.8~MHz, 26~MHz, 27~MHz, 43.2~MHz and 43.4~MHz. When driven axially, the deformation profile due to magnetostriction is also radial, as shown in Figure~\ref{Fig:hv}e. This suggests the magnetometer will perform well when axially driven near radial breathing modes. Choosing axial field magnetic field driving, we find the power spectral density, network response and sensitivity shown in Figure~\ref{Fig:hv}d. The radial breathing mode at $\Omega_{\rm M}/2\pi=27~$MHz, third from left in Figure~\ref{Fig:hv}d top inset, reaches a sensitivity of 180~p$\T$. We confirm that the result from multi-mode analysis (see Figure~\ref{Fig:hv}d bottom blue line) is consistent with single mode analysis (see Figure~\ref{Fig:hv}d bottom triangles) for this mode.  The actuation parameter is 3200 times larger than if the same mechanical mode is driven by an in-plane magnetic field (see Appendix~\ref{app:cact} for $c_{act}$ values), verifying a strong dependence of the magnetomechanical overlap on the magnetic field direction and the potential for vectorial magnetometry.

With in-plane magnetic field driving, the sensitivity observed in the experiment 200~p$\T$ \cite{Forstner2014} surpasses the modelled sensitivity by around two orders of magnitude. This is likely due to the fact that the simulated mode at 23~MHz (Figure~\ref{Fig:hv}b bottom inset) is thermally resolved in the experiment,   which is not the case in the model. This difference can be understood in terms of symmetry. \mbox{In the model}, the~symmetry results in a very poor predicted  optical transduction sensitivity. However, \mbox{in the} experiment, it can be expected that the symmetry is broken due to fabrication defects resulting in improved sensitivity \cite{modal}.

\section{Conclusions}
We have developed a new versatile approach to model the sensitivity of optomechanical magnetometers, introducing magnetostriction into the elastic wave equation used to solve for mechanical eigenmodes. By numerically solving a modified elastic wave equation for a range of geometries, we model the sensitivity for magnetometers both experimentally demonstrated and not-yet fabricated. The modelling predicts that at least one order of magnitude improvement from previous experimental results \cite{Forstner2014} is possible. The sensitivity of \mbox{opto}mechanical magnetometers can be significantly improved by optimising the size and the shape of the \mbox{Terfenol-D}, by utilizing the bending effect, which arises from a magnetic equivalent of the bi-metallic strip effect, and by optimizations of the composition and the annealing process of Terfenol-D, which may lead to sensitivity below 20~p$\T$ using the piezomagnetic constant in Ref.~\cite{Ostenson1990} with micrometre-level resolution.

The numerical method developed here is applicable to optomechanical magnetometers with a wide range of geometries and any magnetostrictive materials. A full characterization of the response of the magnetomechanical overlap to the variation of the signal magnetic fields direction may allow vectorial optomechanical magnetometry, complementary to vectorial optomechanical force sensors \cite{Rossi2016,Arcizet2016}. Micro-optomechanical magnetometers with p$\T$ sensitivity can potentially be applied to detect signals from neurons, similar to recent results with nitrogen-vacancy centre based magnetometers \cite{Barry2016} and atomic magnetometers \cite{Jensen2016}, but with benefits of a simpler, silicon-chip fabricateable approach, as well as high bandwidth.

\section*{Acknowledgements}
We appreciate ETREMA Products, Inc. {(Ames, IA, US)} for providing advice on the choice of Terfenol-D parameters for simulation, and we thank Christopher~Baker, Bei-Bei~Li, George~A.~Brawley, Kiran~E.~Kholsa and James~S.~Bennett for useful discussions. This research is funded by Australian Research Council Discovery Project DP140100734, and Defence Science and Technology Group projects CERA49 and CERA50. W.B. acknowledges the Australian Research Future Fellowship FT140100650.

\section*{}
\appendix

\section{Derivation of the Sensitivity for a Single Mechanical Mode \label{app:single}}
For a single mechanical mode, the minimum detectable magnetic field can be obtained from the actuation parameter $c_{act}$ and the noise force spectral density $S_{FF}$. Calibrated in the medium of air, \mbox{the sensitivity} at individual mechanical eigenfrequencies can be written as

\begin{align}
B_{\rm min}&=\frac{1}{\sqrt{2\pi}c_{act}}\sqrt{S^{\rm therm}_{FF}+S^{\rm imp}_{FF}+S^{\rm ba}_{FF}}\label{BminLTherm},
\end{align}
where a factor of $1/\sqrt{2\pi}$ ensures $B_{\rm min}$ having the unit of $\T$. Noise sources considered are thermal noise and noise from optical measurement including imprecision and back-action. Measurement imprecision comes from the laser shot noise in the optical phase quadrature. Back-action noise is due to the laser shot noise in the optical amplitude quadrature driving the mechanical oscillator. 

The power spectral density in the unit of force (specifically rad$\cdot$s$\cdot$N$^2$) for individual mechanical modes driven by noise can be found in {Ref.}~\cite{Bowen2015}. Here, we extend the calculation to include the back-action noise, as:

\begin{align}
& S^{\rm therm}_{FF}(\Omega_{\rm M})=4 m_{\rm eff}(\Omega_{\rm M})\Gamma(\Omega_{\rm M})k_B T,\label{Stherm}\\
& S^{\rm imp}_{FF}(\Omega_{\rm M})=\frac{m_{\rm eff}(\Omega_{\rm M})\hbar Q_{\rm M}(\Omega_{\rm M})}{8\eta\lvert\chi(\Omega_{\rm M})\rvert^2\Big\lvert C_{\rm eff}(\Omega_{\rm M})\Big\rvert},\label{Simp}\\
& S^{\rm ba}_{FF}(\Omega_{\rm M})=4 m_{\rm eff}(\Omega_{\rm M})\Gamma(\Omega_{\rm M})\hbar\Omega_{\rm M}\lvert C_{\rm eff}(\Omega_{\rm M})\rvert,\label{Sba}
\end{align}
in which a factor of 4 in front of the classical thermal force spectrum in Equation~\eqref{Stherm} is due to the definition of $\Gamma$ being the full-width-half-maximum of the mechanical oscillator. 

Inspection of Equations \eqref{Simp} and ~\eqref{Sba} shows that, despite optomechanical coupling, the effective cooperativity $C_{\rm eff}$ also quantifies the trade-off between better measurement precision and large back-action noise due to the Heisenberg uncertainty relation. $\lvert C_{\rm eff}(\Omega)\rvert$ is given by

\begin{align}
\lvert C_{\rm eff}(\Omega)\rvert \equiv \frac{4g_0^2(\Omega_{\rm M}) N}{\kappa\Gamma\big | 1-\frac{2i\Omega}{\kappa}\big|^2}=\frac{16\eta_{esc} g_0^2(\Omega_{\rm M}) N_{in}}{\Gamma\big|\kappa-2i\Omega\big|^2},\label{ceffdetune}
\end{align}
where $N$ is the intra-cavity photon number,  $N_{in}$~(photons $\cdot$s$^{-1}$) is the input photon number  flux, and~$g_0$~(rad$\cdot$s$^{-1}$) is the vacuum optomechanical coupling rate quantifying the optical resonance frequency shift by the mechanical displacement at zero energy excitation. Fibre-device coupling here is idealized to be lossless where the intra-cavity and end mirror loss due to the scattering and/or absorption of the light is neglected, leaving the optical decay only counted at the front mirror to be $\kappa$ as shown in Figure~\ref{Fig:setup}a and thus making the cavity escape efficiency $\eta_{esc}=\kappa/(\kappa+0)=1$.

To provide an idea of the relative magnitude of $S^{\rm therm}_{FF}$, $S^{\rm imp}_{FF}$ and $S^{\rm ba}_{FF}$, we choose the geometry and parameters (in Section~\ref{multi-simu} in the main text) of the magnetometer reported in Ref.~\cite{Forstner2014}. Not surprisingly, on mechanical resonances at room temperature, we find that back-action noise is always smaller than the thermal noise by a large margin. For each mechanical mode, this is quantified, roughly, by the ratio of thermal phonon occupancy to effective cooperativity, with the former being in the range of $10^5$--$10^6$ for our mechanical frequencies at room temperature, and the later not exceeding 100 for typical parameters. Far from mechanical resonance, the backaction noise will eventually exceed the thermal noise~\cite{Bowen2015}. However, in this regime, the optical shot noise dominates. As a consequence, backaction noise can be safely neglected at all frequencies.

\section{{COMSOL Implementation of the Magnetic Field} \label{app:MagneticField}}
\begin{figure}[!ht]
\begin{center}
\includegraphics[width=0.9 \columnwidth]{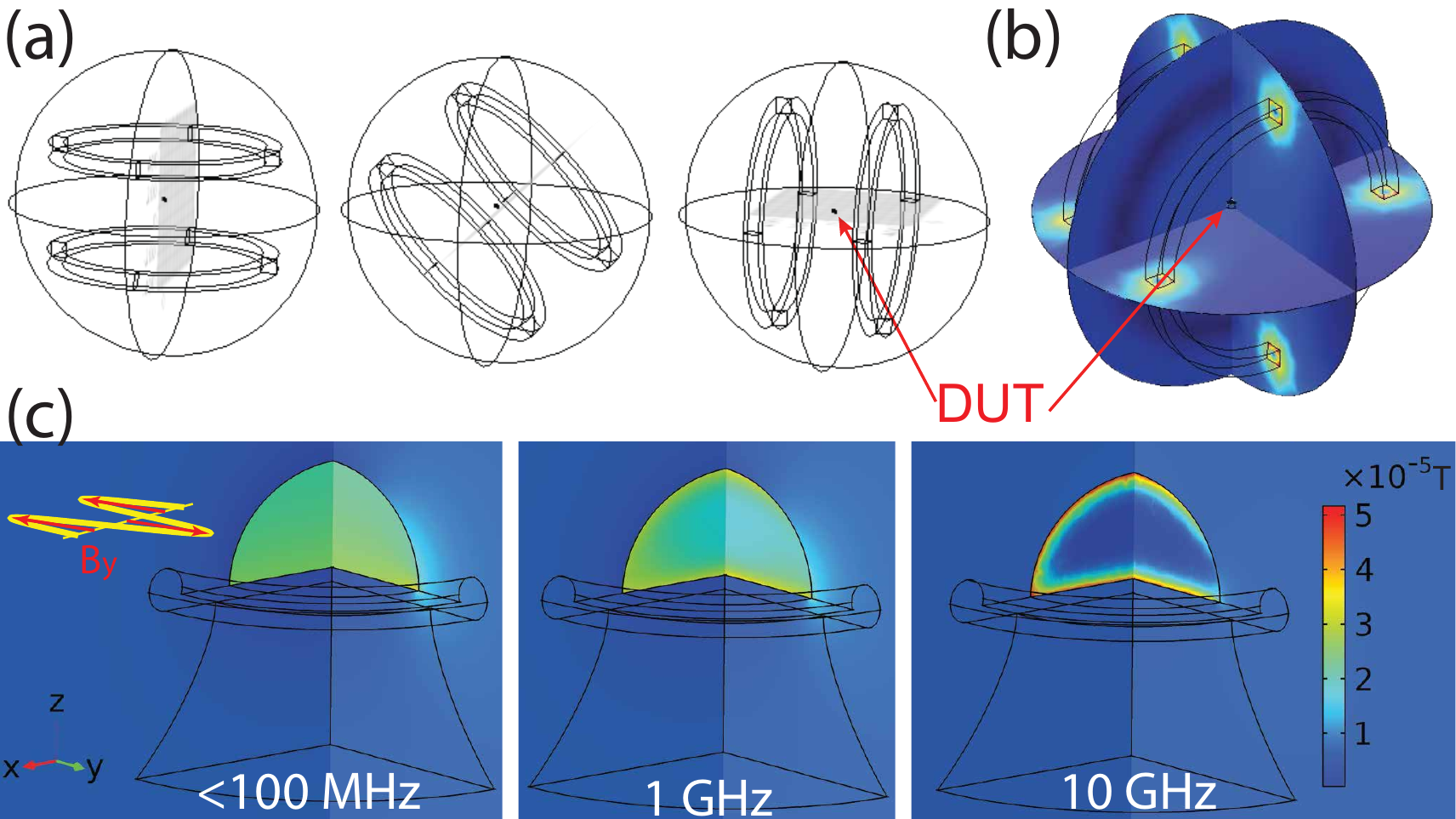}
\end{center}
\vspace{-3mm}\caption{
(\textbf{a}) COMSOL layout for a pair of Helmholtz coils used to generate signal magnetic field. The axis of the pair of coils can be freely rotated in a 4$\pi$ solid angle. The diametre of the coils are more than 40 times larger than the lateral size of the Terfenol-D in the device under test (DUT); \mbox{(\textbf{b}) intersected} orthonormal slices are used to project the amplitude of the magnetic field; (\textbf{c}) the effect of the eddy current inside Terfenol-D when the signal magnetic field $B_{\rm sig}$ is driven in plane with frequency below 100~MHz, at 1~GHz and at 10~GHz. Colourmap refers to the magnetic field inside the Terfenol-D.
}
\label{Fig:eddy}
\end{figure}

The magnetic field $B_{\rm sig}$ is generated by a pair of Helmholtz coils whose axis can be freely rotated in a 4$\pi$ solid angle as shown in Figure~\ref{Fig:eddy}a for COMSOL layout. To enable the simulation of the magnetic field, the outermost sphere is filled with air. The amplitude of the magnetic field is controlled by inputting a known current in the pair of Helmholtz coils. The coils diametre is set to be more than 40 times larger than the lateral size of the Terfenol-D to ensure a uniform driving magnetic field. Therefore, the direction of $B_{\rm sig}$ is along the axial axis of the pair of coils. The magnetic field can be viewed by the intersected orthonormal slices on which the magnetic field amplitude is projected, with the colour refers to the amplitude of the magnetic field as shown in Figure~\ref{Fig:eddy}b.

At high frequencies, the eddy current induced magnetic field opposes the external magnetic field and thus reinforces most of the magnetic field between the surface and skin depth, leaving most inner part of the magnetostrictive material unused. This undesired effect is evaluated by varying the frequency of the signal magnetic field $B_{\rm sig}$. The magnetometer \cite{Forstner2012} as illustrated in Figure~\ref{Fig:singlemodes}a left is used to perform the eddy current simulation. Figure~\ref{Fig:eddy}c shows that the skin depth effect starts to take place at frequencies above 1~GHz. The in-plane magnetic field $B_{\rm sig}$ characterized at the location of Terfenol-D is 7.7~$\upmu$T, and the colourmap shows the magnetic field inside the Terfenol-D is  $\sim25~\upmu$T at low frequencies as shown in Figure~\ref{Fig:eddy}c left. The value from the colourmap is consistent with the calculation from $B_{\rm sig}$ and the value of the relative permeability tensor in Table~\ref{tab:coeffi}. \mbox{The simulation} agrees with the simple relation for skin depth $\delta_{\rm skin}=1/ \sqrt{\pi f \mu_0\mu_r \sigma_c},$ where the conductivity for Terfenol-D is $\sigma_c=$  1.67$\times10^6$ S/m. Mechanical eigenfrequencies of interest of modelled devices are below 50~MHz, far detuned from the range in which eddy currents pose a problem.

\section{COMSOL Implementation of the Modified Elastic Wave Equation \label{app:ElasticW}}
COMSOL's Solid Mechanics module allows users to modify the elastic wave equation and solve it numerically. In the modified elastic wave equation given by Equation~\eqref{EqMdamping} displacement $\bm u=\sum_{i} u_i(\bm r,\Omega)\bm e_i$, and stress tensor $\bm \sigma=\sum_{ij}\sigma_{ij} \bm e_{ij}=\sum_{n}\sigma_{n} \bm e_{n}$ in three dimensions can be fitted into a $3\times 3$ matrix as

\begin{align}
\label{matsig}
\bm \sigma=
\begin{bmatrix}
\sigma_{xx} & \sigma_{xy} & \sigma_{xz} \\
\sigma_{xy} & \sigma_{yy} & \sigma_{yz} \\
\sigma_{xz} & \sigma_{yz}  & \sigma_{zz} 
\end{bmatrix}
=\begin{bmatrix}
\sigma_1 & \sigma_6 & \sigma_5 \\
\sigma_6 & \sigma_2 & \sigma_4 \\
\sigma_5 & \sigma_4  & \sigma_3 
\end{bmatrix}.
\end{align}

Up to the first term in the bracket on the right-hand side (RHS) of Equation~\eqref{EqMdamping} is the elastic stress $\bm \sigma_{\rm ela}$. Elastic stress is connected to strain via a tensor coefficient $\lambda_{ijkl}$ as $\sigma_{\rm ela}^{ij}=\lambda_{ijkl}\epsilon_{kl}$. The tensor with $\lambda_{ijkl}$ being its elements is termed {\textit{elasticity matrix}} in COMSOL. For isotropic materials, elements of the elasticity matrix are determined by an isotropic Young's modulus and an isotropic Poisson's ratio, while for anisotropic materials the number of independent elements can go up to 21 in the \mbox{$6\times6$ matrix \cite{LandauElas}}. Terms  $\bm \sigma_{\rm ela}+\bm \sigma_{\rm driv}$ in the bracket on the RHS of Equation~\eqref{EqMdamping} incorporate both the elastic stress and the stress under the magnetic field driving. Assuming (1) the variation of the AC magnetic field is slow enough for the material to reach deformed equilibrium, (2) the magnetostrictive material exhibits reversibility, and (3) the operational point is far below the magnetostrictive saturated strain defined as the ratio of the maximum material elongation to its original length, the stress-magnetic field relation can be linearly approximated \cite{Terfenol}. $\bm \sigma_{\rm ela}+\bm \sigma_{\rm driv}$ with small modulation can be expressed via first order Taylor expansion, when projected onto one dimension, as

\begin{equation}
\begin{aligned}
&\Delta\sigma^{ij}_{\rm ela}+\Delta\sigma^{ij}_{\rm driv}=\underbrace{\frac{\partial\sigma^{ij}}{\partial  \epsilon_{kl}}\bigg|_{\rm H} \Delta\epsilon_{kl}+\mathcal{O}(\Delta\epsilon_{kl}^2)}_{\Delta \sigma^{ij}_{ela}}\\
&+\underbrace{\frac{\partial  \sigma^{ij}}{\partial\mathrm{H}_{k}}\bigg|_\epsilon \Delta \mathrm{H}_{k}+\mathcal{O}(\Delta\mathrm{H}_{k}^2)+\Delta \sigma^{ij}_{\rm Maxw}}_{\Delta \sigma^{ij}_{\rm driv}}\\
&\approx\lambda_{ijkl}\super H\Delta \epsilon_{kl}+e\superl \epsilon_{ijk} \Delta\mathrm{H}_{k}+\Delta\sigma^{ij}_{\rm Maxw},
\label{stressconsti}
\end{aligned}
\end{equation}
where the modulated Maxwell stress tensor $\Delta \sigma^{ij}_{\rm Maxw}$ ~\cite{GriffithsEM} {describes} the stress caused by the interaction of a magnetized ferromagnetic material and the external magnetic field, and its value depends on the relative permeability of the magnetostrictive material. $\mathrm{H}_{k}$ is the magnetic field strength inside the Terfenol-D when the external magnetic field strength $\mathrm{H}_0=B_{\rm sig}/\mu_0$ ($\mu_0$ is the vacuum permeability) magnetizes the magnetostrictive material due to the effect of the demagnetization field. The internal magnetic field $\mathrm{H}_{k}$ decreases as the length of the magnetic rod is reduced from infinite length where $\mathrm{H}_k=\mathrm{H}_0$ to zero-thickness where $\mathrm{H}_k=\mathrm{H}_0/\mu_r$, in which $\mu_r$ is the relative permeability of the magnetostrictive material. 

The last term in the bracket on the RHS of Equation~\eqref{EqMdamping} is the input damping ${\bm \sigma}_\Gamma$. It is chosen to be proportional to the time derivative of strain as $\bm \sigma_{\Gamma}=\eta_{\rm damp}\dot{\bm \epsilon}$ (the unit of $\eta_{\rm damp}$ is Pa$\cdot$s) so that the integral form of the elastic wave equation  along one dimension 

\begin{align}
\int_V \rho_n \frac{\partial ^2  u_i}{\partial t^2}dV=\int_S ( \sigma^{ij}_{\rm ela} + \sigma^{ij}_{\rm driv} +\sigma^{ij}_{\Gamma})\cdot dS
\end{align}
followed by a Fourier transformation to the frequency domain can have the damping coefficient $\Gamma$ in front of the term linear with $\Omega$ in the mechanical susceptibility. Only the part of $( \sigma^{ij}_{\rm ela} + \sigma^{ij}_{\rm driv} +\sigma^{ij}_{\Gamma})$ perpendicular to $dS$ contributes to the integral. Roughly speaking (simplified to one dimension), $m_{\rm eff}\Omega^2_{\rm M}$ is related to elasticity matrix and spatial profile of the mechanical eigenmode {$\Psi(r)$}~and surface area $S$. The damping factor $\Gamma$ includes the input damping $\eta_{\rm damp}$~(Pa$\cdot$s), $m_{\rm eff}$, {$\Psi(r)$} and surface area $S$. The actuation parameter is determined by $c_{act}=e^\epsilon S_\perp/\mu_0\mu_r$ ($S_\perp$ being the area perpendicular to the magnetic field induced stress). Note that {$\Psi(r)$} comes from an ansatz of Equation~\eqref{EqMdamping} as the displacement~ {$u(r,\Omega)= \Psi(r)x(\Omega)$} can be decoupled into a spatial and a frequency dependent term.

Due to the transverse (axial) symmetry of the Terfenol-D, the independent elements of the elasticity matrix are reduced to only six \cite{Terfenol}, which are $\lambda^{\rm H}_{11}$, $\lambda^{\rm H}_{12}$, $\lambda^{\rm H}_{13}$, $\lambda^{\rm H}_{33}$, $\lambda^{\rm H}_{44}$ and $\lambda^{\rm H}_{66}$. The explicit tensor form of stress, strain and magnetic field relation is presented as: 

\begin{align}
\label{constiMatr}
\begin{split}
\begin{bmatrix}
\sigma^{xx}_{\rm total} \\ \vspace{0.5mm}
\sigma^{yy}_{\rm total} \\ \vspace{0.5mm}
\sigma^{zz}_{\rm total} \\ \vspace{0.5mm}
\sigma^{yz}_{\rm total} \\ \vspace{0.5mm}
\sigma^{xz}_{\rm total} \\ \vspace{0.5mm}
\sigma^{xy}_{\rm total}
\end{bmatrix}=
\underbrace{\begin{bmatrix}
\lambda^{\rm H}_{11} & \lambda^{\rm H}_{12} & \lambda^{\rm H}_{13} & 0 & 0 & 0 \\
\lambda^{\rm H}_{12} & \lambda^{\rm H}_{11} & \lambda^{\rm H}_{13} & 0 & 0 & 0 \\
\lambda^{\rm H}_{13} & \lambda^{\rm H}_{13} & \lambda^{\rm H}_{33} & 0 & 0 & 0 \\
0 & 0 & 0 & \lambda^{\rm H}_{44} & 0 & 0 \\
0 & 0 & 0 & 0 & \lambda^{\rm H}_{44} & 0 \\
0 & 0 & 0 & 0 & 0 & \lambda^{\rm H}_{66} 
\end{bmatrix}}_{\text{elasticity matrix}}
\begin{bmatrix}
\epsilon_{xx} \\ \vspace{0.5mm}
\epsilon_{yy} \\ \vspace{0.5mm}
\epsilon_{zz} \\ \vspace{0.5mm}
\epsilon_{yz} \\ \vspace{0.5mm}
\epsilon_{xz} \\ \vspace{0.5mm}
\epsilon_{xy}
\end{bmatrix}+\\
\underbrace{\begin{bmatrix}
\sigma^{xx}_{\Gamma} \\ \vspace{0.5mm}
\sigma^{yy}_{\Gamma} \\ \vspace{0.5mm}
\sigma^{zz}_{\Gamma} \\ \vspace{0.5mm}
\sigma^{yz}_{\Gamma} \\ \vspace{0.5mm}
\sigma^{xz}_{\Gamma} \\ \vspace{0.5mm}
\sigma^{xy}_{\Gamma}
\end{bmatrix}}_{\rm damping}
+ \underbrace{\begin{bmatrix}
0 & 0 & e^\epsilon_{13} \\
0 & 0 & e^\epsilon_{13} \\
0 & 0 & e^\epsilon_{33} \\
0 & e^\epsilon_{15} & 0 \\
e^\epsilon_{15} & 0 & 0 \\
0 & 0 & 0
\end{bmatrix}
\begin{bmatrix}
H_x \\
H_y \\
H_z
\end{bmatrix}}_{\text{external driving stress}},
\end{split}
\end{align}
where small modulation $\Delta \epsilon_{kl}$ and $\Delta\mathrm{H}_{k}$ in Equation~\eqref{stressconsti} is replaced with $\epsilon_{kl}$ and $H_{k}$ in the frequency domain. The input elasticity matrix elements $\lambda\super H_{ijkl}$ and piezomagnetic constant elements $e^\epsilon _{ijk}$ are taken from an experimental measurement biased at 60~kA/m and prestressed at 20~MPa \cite{Claeyssen1991} as summarised in Table~\ref{tab:coeffi}. In addition, the density of Terfenol-D  and silica are taken as 9250~kg$\cdot$m$^{-3}$ and 2203~kg$\cdot$m$^{-3}$, respectively. The Maxwell stress tensor in Equation~\eqref{stressconsti} is neglected in Equation~\eqref{constiMatr}. This is because the contribution of the Maxwell stress tensor and magnetostrictive stress (taking the value from Table~\ref{tab:coeffi}) is comparable only when the driving magnetic field is no less than the order of 10 Tesla under linear stress-magnetic-field approximation. Due to the large piezomagnetic constant of magnetostrictive materials, the influence from the Maxwell stress tensor in the Terfenol-D can be safely neglected in experimental condition where magnetic field is well below microtesla (magnetostrictive stress is $\sim 10^6$ times larger than the Maxwell stress). The input external driving stress from Equation~\eqref{constiMatr} is fitted into a 3$\times$3 matrix in the form of Equation~\eqref{matsig} in COMSOL as external stress after simple matrix product calculation.

\begin{table}[!ht]
\centering
\caption{ Coefficients in the magnetomechanical coupling biased at 60~kA/m and prestressed at 20~MPa \cite{Claeyssen1991}. $\lambda^{\rm H}$ is the elasticity matrix element, $e^{\epsilon}$ is the piezomagnetic constant and $\mu^\sigma$ is the relative magnetic permeability.}
\begin{ruledtabular}
\begin{tabular}{c c c c c c c c} 
\toprule
unit (GPa) & $\lambda^{\rm H}_{11}$ & $\lambda^{\rm H}_{12}$ & $\lambda^{\rm H}_{13}$ & $\lambda^{\rm H}_{33}$ & $\lambda^{\rm H}_{44}$ & $\lambda^{\rm H}_{66}$\\ 
 & 107 & 74.8 & 82.1 & 98.1 & 60 & 161\\
\hline
unit (T) & $e^{\epsilon}_{13}$ & $e^{\epsilon}_{33}$ &  $e^{\epsilon}_{15}$ & no unit & $\mu^{\sigma}_{11}$ & $\mu^{\rm \sigma}_{33}$\\ 
 & 90 & $-$166 & $-$168 & & 6.9$\mu_0$  & 4.4 $\mu_0$ \\
\bottomrule
\end{tabular}
\label{tab:coeffi}
\end{ruledtabular}
\end{table}

\section{Calculation of $c_{act}$ by Lorentzian Fit \label{app:cact}}
Actuation parameter $c_{act}$ is extracted by fitting the equation of motion to the Lorentzian distribution. In COMSOL, the phase of the mechanical displacement spectrum is in the range 0 to $-\pi/2$  at frequencies below the resonance frequency and $\pi/2$ to 0 above it, this generates an artefact that we remove by taking the absolute value of the displacement. The modified mechanical frequency response given by the Fourier transform of Equation~\eqref{asd} for each single mechanical mode is given by

\begin{align}
\frac{max[\lvert \bm u  \rvert]\cdot  m_{\rm eff}}{\lvert B_{\rm sig}\rvert}=\frac{c_{act}}{\lvert -\Omega^2-i\Gamma\Omega+\Omega_{\rm M}^2\rvert}.\label{fitEq}
\end{align}

All the fitting parameters are on the RHS of Equation~\eqref{fitEq}: $c_{act}$, $ \Gamma$ and $\Omega_{\rm M}$, while all the three parameters on the LHS of Equation~\eqref{fitEq}, maximum displacement, effective mass and signal magnetic field, can be drawn from COMSOL. The COMSOL syntax for extracting the maximum displacement is~{\texttt{sqrt(abs(u\string^2)+abs(v\string^2)+abs(w\string^2))}} under volume maximum analysis and the $m_{\rm eff}$ is the quotient of {\texttt{solid.rho*(abs(w\string^2)+abs(v\string^2)+abs(u\string^2))}} under volume integration analysis and maximum displacement where {\texttt{u,v,w}} are displacements in $x,y,z$ directions.

\begin{figure}[!ht]
\begin{center}
\includegraphics[width=0.9 \columnwidth]{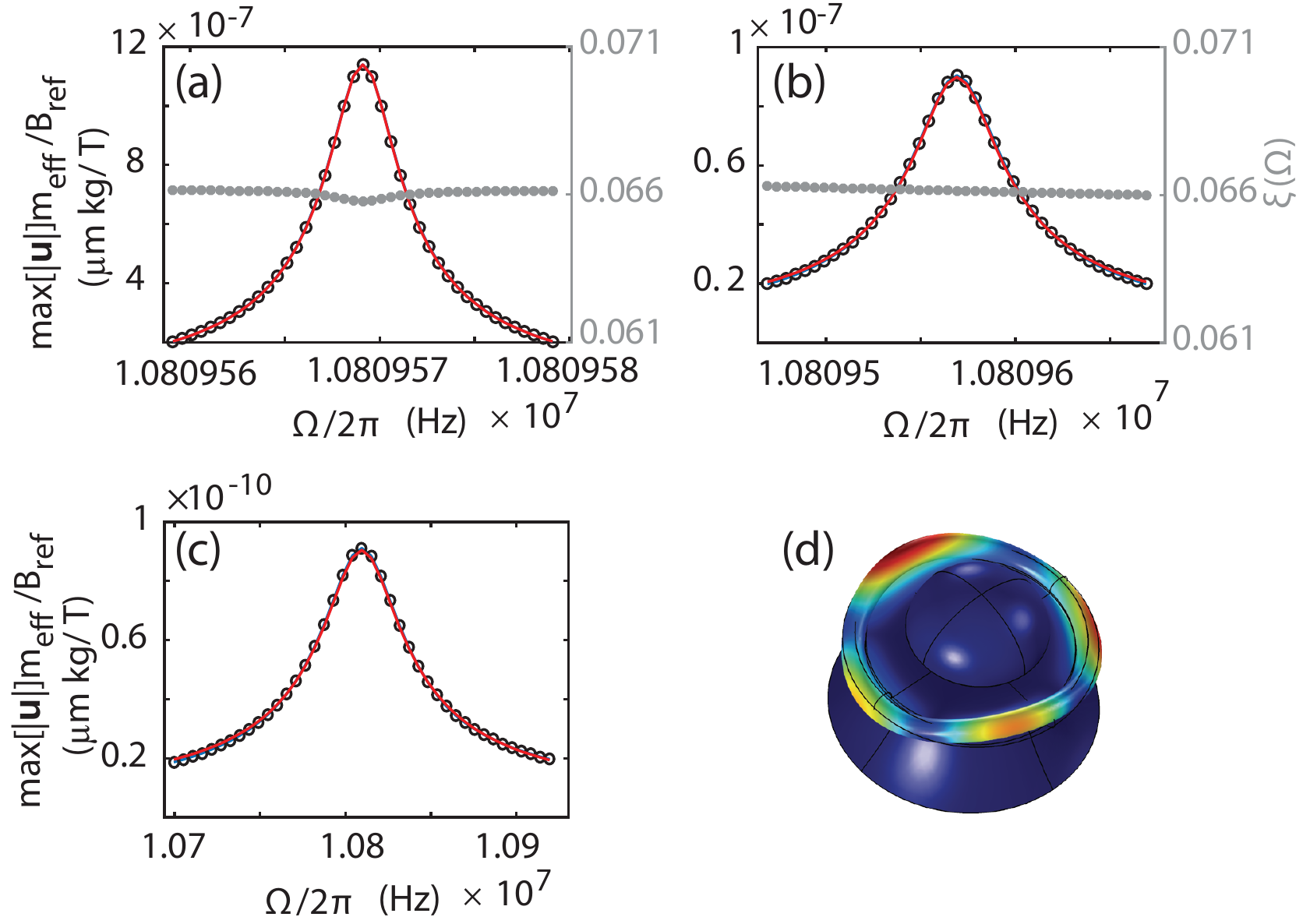}
\end{center}
\vspace{-5mm}\caption{
(\textbf{a}) a fit to the Lorentzian distribution, and $\xi$ spectrum for a small input material damping; (\textbf{b}) the fit and $\xi$ spectrum with input damping factor 12.5 times larger than that of a); (\textbf{c}) the fit with damping factor 12500 times larger than that of a); (\textbf{d}) all the fits and $\xi$ spectrum are performed on the mechanical mode with Terfenol-D position offset from the centre.
}
\label{Fig:fits}
\end{figure}

Since damping is input manually, it is important to check whether the damping affects $c_{act}$ or not. By changing the input damping for a wide parameter range of 12500 times variation, it has been verified that the effect of the damping variation on $c_{act}$ is negligible. Figure~\ref{Fig:fits}a shows the fit to the Lorentzian distribution for an input damping factor 12.5 times smaller than that of in Figure~\ref{Fig:fits}b, and~the fit in Figure~\ref{Fig:fits}c has the damping factor 12500 times larger than that of in Figure~\ref{Fig:fits}a. The fitting is performed on a mechanical mode simulated in Figure~\ref{Fig:fits}d. A smaller input damping allows for a clearer Lorentzian fit as shown in Figure~\ref{Fig:fits}a. Therefore, in the implementation, input damping is chosen to be as small as possible limited by COMSOL's convergence error.

\begin{figure}[!ht]
\begin{center}
\includegraphics[width=0.9 \columnwidth]{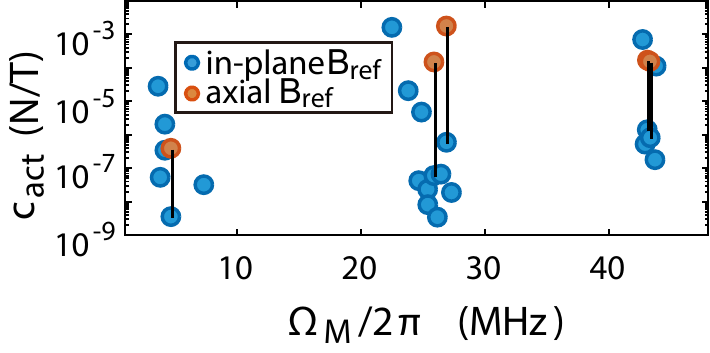}
\end{center}
\vspace{-5mm}\caption{
Actuation parameter $c_{act}$ for the magnetometer \cite{Forstner2014} using the parameters in Section~\ref{multi-simu}. \mbox{A black} vertical line connects the same mechanical mode under magnetic field driving in both in-plane and axial directions. 
}
\label{Fig:cact}
\end{figure}

Figure~\ref{Fig:cact} shows the actuation parameter $c_{act}$ of the magnetometer \cite{Forstner2014} using the parameters in Section~\ref{multi-simu} in three frequency sections of mechanical modes. Blue dots present the $c_{act}$ for mechanical modes under in-plane magnetic field driving, while orange dots are for mechanical modes under axial driving. As can be seen, the actuation parameter varies over many orders of magnitude both for different mechanical modes with the same driving magnetic field direction (comparison of dots among the same colour) and for the same mode driving by the magnetic field in two orthogonal directions (in~comparison between red and blue dots at the same mechanical eigenfrequency, connected via a black vertical line in Figure~\ref{Fig:cact}). 

\section{Calculation of Geometrical Factor $\xi$ \label{app:xi}}
This numerical modelling offers an accurate way to compute the effective cooperativity $C_{\rm eff}$ from the geometrical factor $\xi$ defined in Equation~\eqref{xi}. The optical path length $L$ corresponds to the outermost circumference in micro-toroidal resonators. The change of circumference $\delta L(\Omega)$ in a harmonic oscillation at a mechanical resonance can be obtained by synchronizing the output displacement to the maximum and minimum amplitudes. The implementation of the synchronization in COMSOL is to multiply a phase shift in all three Cartesian coordinates in the deformable mesh setting as

\begin{equation}
\begin{aligned}
\texttt{dx=u*exp(-i*atan(imag(u)/(real(u)+1e-16)))},\\
\texttt{dy=v*exp(-i*atan(imag(v)/(real(v)+1e-16)))},\\
\texttt{dz=w*exp(-i*atan(imag(w)/(real(u)+1e-16)))},
\label{uvwsyn}
\end{aligned}
\end{equation}
where {\texttt{1e-16}} in the denominator is an example of adding a small value to eliminate the error of dividing by 0 when {\texttt{real(u)/real(v)/real(w)}} is 0, and {\texttt{dx,dy,dz}} are  displacements after the synchronization. 

A line integral with integrand 1 along the outermost circumference of the devices is followed after the phase synchronization, which results in the circumference at a particular phase. Linear optomechanical coupling is evaluated through $\delta L$ by taking the difference of circumference synchronized at phase 0 and phase $\pi$ where Equation~\eqref{uvwsyn} is multiplied by a phase factor {\texttt{exp(i*pi)}}. Figure~\ref{Fig:fits}a,b shows the constant $\xi$ across a mechanical resonance with 12.5 times variation of input damping, showing that $\xi$ is insensitive to the mechanical quality. $\xi$ is missing in Figure~\ref{Fig:fits}c due to the numerical error at large manually input damping where $\xi$ spectrum is far away from constant. Plotting $\xi$ spectrum offers a way of sanity check of possible numerical errors.

\section{From $\xi$ to Optomechanical Coupling \label{app:fromxi}}
With the calculated geometrical factor $\xi$, the value of the parameters describing optomechanical coupling $C_{\rm eff}$ is easy to achieve. The optomechanical coupling rate is defined as $g_0\equiv G\cdot x_{\rm zpf}$, where zero-point motion $x_{\rm zpf}=\sqrt{\hbar/(2m_{\rm eff}\Omega_{\rm M})}$, and $G$~(rad$\cdot$m$^{-1}$s$^{-1}$) is the optomechanical coupling strength quantifying the shift of optical resonance frequency $\delta\omega_0$ by the mechanical displacement as 

\begin{align}
G=\frac{\delta\omega_0}{max[\bm u] }.\label{G}
\end{align}

For a Fabry--P\'erot type and  micro-toroidal structure cavity with length $L$, the shift of the optical resonance frequency is linked with the change of the cavity length by $\delta L/L=\delta\omega_0/\omega_0$. Inserting the expression $\delta L/L=\delta\omega_0/\omega_0$ into Equation~\eqref{G} leads to 

\begin{align}
G=\omega_0 \cdot\xi/L.
\end{align}

Therefore, $g_0$ and thereby $C_{\rm eff}$ can then be written as a function of the geometrical factor $\xi$. For an individual mechanical eigenmode at $\Omega_{\rm M}$ and based on Equation~\eqref{ceffdetune}, the expression for the effective cooperativity  as a function of the geometrical factor $\xi$ can be calculated to

\begin{equation}
\begin{aligned}
\lvert  C_{\rm eff}(\Omega)\rvert=\frac{\xi(\Omega_{\rm M})^2}{m_{\rm eff}(\Omega_{\rm M})\Omega_{\rm M}\Gamma(\Omega_{\rm M})}\cdot\frac{8\eta_{esc}\hbar N_{in}\omega_0^2}{L^2\big( \kappa^2+4\Omega^2\big)},\label{ceffdetun}
\end{aligned}
\end{equation}
where the front part on the RHS is mode dependent, $\xi(\Omega_{\rm M})$ and $m_{\rm eff}(\Omega_{\rm M})$ can be accurately extracted from the numerical simulation. The optical resonance frequency $\omega_0$ and length of the cavity $L$ are input parameters, and the only empirical parameters left are the optical decay $\kappa$ and the mechanical damping factor $\Gamma$ with the assumption of an idealized lossless cavity escape efficiency $\eta_{esc}=1$.

\end{document}